\documentstyle[12pt]{article}

\newcommand{\beq}{\begin{equation}}
\newcommand{\eeq}{\end{equation}}
\newcommand{\beqa}{\begin{eqnarray}}
\newcommand{\eeqa}{\end{eqnarray}}
\newcommand{\nn}{\nonumber}
\newcommand{\eq}[1]{(\ref{#1})}

\newcommand\delt{\partial_{\theta}}
\newcommand\delx{\partial_x}          
\newcommand{\NP}[1]{ {\it Nucl.~Phys.} {\bf #1}}
\newcommand{\PL}[1]{ {\it Phys.~Lett.} {\bf #1}}

\newcommand{\PTP}[1]{ {\it Prog.~Theor.~Phys.} {\bf #1}}

\newcommand{\MPL}[1]{ {\it Mod.~Phys.~Lett.} {\bf #1}}
\newcommand{\IJMP}[1]{ {\it Int.~Jour.~Mod.~Phys.} {\bf #1}}
\newcommand{\JP}[1]{ {\it J.~Phys.} {\bf #1}:\  Math.~Gen.~}
\begin{document}
\topmargin 0pt
\oddsidemargin 1mm
\begin{titlepage}
\begin{flushright}
NBI-HE-97-34\\
hep-th/9709047\\
\end{flushright}
\setcounter{page}{0}
\vspace{15mm}
\begin{center}
{\Large Deformation of Super Virasoro Algebra in Noncommutative 
Quantum Superspace} 
\vspace{20mm}

{\large Haru-Tada Sato   
\footnote{sato@nbi.dk}}\\
{\em The Niels Bohr Institute, University of Copenhagen\\
     Blegdamsvej 17, DK-2100 Copenhagen, Denmark}\\
\end{center}
\vspace{7mm}

\begin{abstract}
We present a twisted commutator deformation for $N=1,2$ super 
Virasoro algebras based on $GL_q(1,1)$ covariant noncommutative 
superspace. 
\end{abstract}

\vspace{1cm}
\vfill
\begin{flushleft}
\end{flushleft}

\end{titlepage}
\newpage
\renewcommand{\thefootnote}{\arabic{footnote}}
%
Several years ago, a deformation of Virasoro algebra was studied 
based on a deformed Heisenberg oscillator or a deformed differential 
operator. The deformation comprises a particular type of twisted 
commutators, firstly proposed by Curtright and Zachos \cite{CZ}, and 
it so far has nothing to do with a Hopf algebra structure, which is 
an important notion for quantum deformation. In this sense, this 
deformation differs from ordinary quantum deformation, however is an 
algebra of well-defined differential operators; i.e., oscillators. 
We refer to this algebra as {\it CZ algebra} in this paper, and a 
review can be found in \cite{history} (where a connection to other 
deformation \cite{V3} is discussed as well).

The physical importance of CZ algebra seems to lie in the following 
regime. The idea of deformed Heisenberg algebra may be useful 
for constructing various objects in an algebraic manner, and would 
be fundamental to discuss a quantum excitation of deformed 
fluctuating vacuum. Thus a deformed symmetry algebra of Virasoro 
type should naturally be expressed in terms of a deformed oscillator 
algebra. {}~For example, it is found that a $W$-algebra extension 
\cite{WCZ} of CZ algebra emerges from a finite (deformed oscillator) 
representation of $c=1$ matrix model \cite{mat}. This is the example 
of physical implication of CZ algebra in a fluctuating system.

The generators of CZ algebra can also be constructed on a quantum 
group basis \cite{hasCZ} covariant under a matrix action, where 
matrices form a (quantum) group and their matrix {\it entries} are 
noncommutative. Accordingly, any component of the basis can 
generally be a noncommutative quantity and become suitable to 
formulate such a fluctuating object where the notion of 
commutable point may not be allowed. We shall call this 
{\it noncommutative space} ``quantum space''.

Our intension is to generalize this approach into a supersymmetric 
case introducing a basis covariant under a super quantum group action. 
The purpose of this letter is to present a simple supersymmetric 
extension of the CZ algebra, allowing $U(1)$ element's participation. 
This is a concise deformation compared to the previous awkward 
results \cite{AKS}, and is nontrivial since there is no 
well-established method to construct a super CZ algebra.

In \cite{AKS}, we discussed a method how to extend the deformed Virasoro 
(CZ) algebra into a superalgebra. The bosonic elements of the deformed 
algebra satisfy 
\beqa
&&[F_n,F_m]_{(m-n)}=0 \ , \label{algFF} \\
&&[L_n^{CZ},F_m]_{(m-n)}=-q^{-n}[m]F_{n+m}\ , \\
&&[L_n^{CZ},L_m^{CZ}]_{(m-n)}=[n-m]L_{n+m}^{CZ} \label{algCZ}\ ,
\eeqa
where $[x]=(q^x-q^{-x})/(q-q^{-1})$ and twisted (anti-)commutators 
are defined as 
\beq
[A,B\}_{(a)} = q^a AB \pm q^{-a} BA \ .
\eeq
The $F_n$ is an element of a deformed $U(1)$ algebra \cite{hasKM}, and 
$L_n^{CZ}$ represents the deformed Virasoro algebra (CZ algebra). 
These operators can be defined in terms of bosonic quantum space 
operators \cite{hasCZ}. For the purpose of supersymmetric extension, 
these operators should be defined on a (noncommutative) quantum 
superspace, and we make use of the following $GL_q(1,1)$ covariant 
quantum superspace \cite{QS} 
\beqa
&&(\theta)^2=(\partial_\theta)^2=0, \quad x \theta =  q\theta x,
\quad \delx \delt =  q^{-1}\delt \delx,        \nn   \\
&&\delx x=1+q^{-2} x \delx, \quad \delt \theta 
           = 1- \theta \delt +(q^{-2}-1)x \delx, \label{qspace} \\
&&\delx\theta=q^{-1}\theta\delx, \quad \delt x=q^{-1}x\delt, \nn
\eeqa
where $x$ and $\theta$ are the bosonic and fermionic (noncommutative) 
coordinates. In the previous paper, we introduced the operator 
($\mu=\delx x -x\delx$) which commutes with $\theta$ and $\delt$ 
as a scaling operator, however this time, we consider the 
scaling operator such that (for a real number $a$) 
\beqa
&&\lambda x=q^{-2}x \lambda, \qquad 
  \lambda \delx =q^2 \delx\lambda,    \label{lamx}\\
&&\lambda\theta=q^{2a}\theta\lambda\qquad  
\lambda\delt=q^{-2a}\delt\lambda\ .         \label{lamth}
\eeqa  
The $\lambda$ operator is realized by 
\beq
   \lambda = \mu+(q^{2a}-1)\theta\delt \ . 
\eeq

According to the program \cite{AKS} to construct a super CZ algebra, 
we first define the $G_r$ operator
\beq
G_r = \lambda^{-1/2}x^{r+{1\over2}}(\delt-\theta\delx)\ .
\eeq
Here we fix $a=-1/2$ imposing the relation 
\beq
       \lambda G_r = q^{-2r}G_r\lambda \ ,
\eeq
and we thus reckon 
\beq
G_rG_s = -q^{r+s+3/2}x^{r+s+1}\delx 
   - q^{r+2}[s+{1\over2}]x^{r+s}\theta\delt\ . \label{lbyg}
\eeq
Remarkably, the following twisted anti-commutator can be cast into 
the form which depends only on $r+s$ (up to an overall factor)
\beq
\{G_r,G_s\}_{({s-r\over2})} = 
   - q^{5/2+r+s} (q^{s-r\over2}+q^{r-s\over2})
(q^{-1}x^{r+s+1}\delx + {1-q^{-r-s-1}\over q-q^{-1}} x^{r+s}\theta\delt)\ .
\label{grgs}
\eeq
One may express RHS of Eq.\eq{grgs} by two bases, for example 
$L_n^{CZ}$ and $F_n$, as discussed in \cite{AKS}, {\it however} we 
here simply propose the bosonic operator $L_n$ defined in terms 
of the RHS; namely 
\beq
L_n = -q^{-1}x^{n+1}\delx - A_n x^n\theta\delt, \qquad 
A_n={1-q^{-n-1}\over q-q^{-1}} \ .
\eeq
This $L_n$ is different from the $L_n^{CZ}$ operator 
\beq
L_n^{CZ} = -q^{-1}x^{n+1}\delx - A_n^{CZ} x^n\theta\delt, \qquad 
A_n^{CZ} = c_1 q^{-2n} + c_2 \ ,
\eeq
where $c_i$ ($i=1,2$) are arbitrary constants, 
and the relation between these two Virasoro-like operators is 
\beq
L_n^{CZ}= L_n +\epsilon_n F_n, \qquad \epsilon_n = A_n-A_n^{CZ},
\eeq
where the deformed $U(1)$ operator appears 
\beq
F_n = x^n\theta\delt \ .
\eeq

We therefore find the algebras 
among $L_n$ and $F_n$ besides the superalgebra of $G_r$;
\beq
\{G_r,G_s\}_{({s-r\over2})} = 
              q^{5/2+r+s} (q^{s-r\over2}+q^{r-s\over2})L_{r+s},
\label{algGG}
\eeq
\beq
[L_n, G_r]_{(r-{n\over2})} = q^{-n}[{n\over2}-r] G_{n+r},
\label{algGL}
\eeq      
\beqa
&&[L_n,F_m]_{(m-n)}= -q^{-n}[m]F_{n+m}\ , \\
&&[L_n,L_m]_{(m-n)}= [n-m]L_{n+m} + a_{nm}F_{n+m} \ ,
\label{algLL}
\eeqa
where 
\beq
a_{nm}= q^{-n-m-1}(q-q^{-1})[{n\over2}][{m\over2}][{n-m\over2}]\ .
\eeq
As expected, $L_n$ is not an independent operator since we have the 
relation from Eq.\eq{lbyg}
\beq
L_n = q^{-5/2-n} G_{n/2}^2 \ .  \label{newLn}
\eeq 
Note that $\lambda$ is related to $L_0$ by 
\beq
\lambda = 1+(q-q^{-1})L_0\ , 
\eeq
and also that $\lambda$ plays the role of a central element 
\beq
[\lambda,E_n]_{(n)} =0 \ ,
\eeq
where $E_n$ stands for an arbitrary element of $G_n$, $L_n$ or $F_n$. 
This deformed superalgebra contains the following $q$-$su(1,1)$ 
and $q$-$osp(1,2)$ algebras of the CZ-type deformation: 
For the $su(1,1)$ part, (noticing $a_{nm}=0$ for $n,m=0,\pm1$), 
we have  
\beq
[L_n,L_m]_{(m-n)}= [n-m]L_{n+m}\ , \qquad (n,m=0,\pm1),
\eeq
and for the $osp(1,2)$ part ($r=\pm1/2$), we have 
\beqa
&& \{G_r,G_{-r}\}_{(-r)} = 
      q^{5/2} (q^r + q^{-r})L_0\ , \qquad 
(G_{\pm 1/2})^2 = q^{5/2\pm1}L_{\pm1}\ , \\
&&[L_0,G_r]_{(r)}=-[r]G_r\ .
\eeqa

Several remarks are now in order. {}~First one is on the Jacobi 
identity constraints \cite{jacobi};
\beq
[G_t,\{G_r,G_s\}_{({s-r\over2})}]_{({r+s\over2}-t)} 
+ \mbox{c.p.} = 0 \ , \label{jacobGGG}
\eeq
\beq
[F_n,[F_m,F_l]_{(l-m)}]_{(m+l-2n)} + \mbox{c.p.}=0\ ,
\label{jacobFFF}
\eeq
\beq
[L_n,[F_m,F_l]_{(l-m)}]_{(m+l-2n)} + \mbox{c.p.}=0\ ,
\label{jacobLFF}
\eeq
\beqa
&&[L_t,\{G_r,G_s\}_{({s-r\over2})}]_{(r+s-t)}
-\{G_s,[L_t,G_r]_{(r-{t\over2})}\}_
{({r+t-3s\over2})}  \nn\\
&&\hskip 30pt+\{G_r,[G_s,L_t]_{({t\over2}-s)}\}_
{({s+t-3r\over2})} =0 \ , \label{jacobLGG}
\eeqa
and so forth, where c.p. means the cyclic permutations as usual. 
The first two relations \eq{jacobGGG},\eq{jacobFFF} are satisfied 
by Eqs.\eq{algGG},\eq{algGL} and \eq{algFF}. On the other hand, 
the latter two relations give rise to the following constraint 
equations, which are similar to the (bosonic) CZ algebra case 
\cite{HS}
\beq
[m][F_l,F_{n+m}]_{(n+m)}=[l][F_m,F_{l+n}]_{(l+n)} \ ,
\eeq
\beqa
&&q^{5/2+r+s}(q^{r-s\over2}+q^{s-r\over2})[L_t,L_{r+s}]_{({r+s\over2}-t)}
-q^{-t}[{t\over2}-r]\{G_s,G_{t+r}\}_{({r+t-3s\over2})}\nn\\
&&\hskip30pt -q^{-t}[{t\over2}-s]\{G_r,G_{t+s}\}_
{({s+t-3r\over2})} = 0 \ .
\eeqa
The $q$-$su(1,1)$ and $q$-$osp(1,2)$ parts do not give rise to any 
constraint. These constraints should be satisfied when an explicit 
realization is applied, and ensure the associativity of operator 
algebra. For example, those are of course satisfied when the differential 
operators shown above are substituted.  According to this feature, 
we may transform the constraints into algebra forms. 
The constraint algebras, with Eq.\eq{newLn}, are thus 
\beq
F_nF_m = q^{3n+m}\lambda F_{n+m} \ ,
\eeq
\beq
\{G_r,G_s\}_{({3s-r\over2})}=q^{5/2+r+s}(q^{3s-r\over2}
+q^{r-3s\over2})L_{r+s}+
q^{3/2}(q-q^{-1})[r][{s-r\over2}]F_{r+s}\ . \label{conalg}
\eeq
We show two examples of the usage of these constraint algebras; (i) 
One should note that Eq.\eq{algLL} can be checked by using 
Eqs.\eq{newLn}, \eq{algGG},\eq{algGL} and \eq{conalg} in order
\beqa
&&\hskip-35pt[L_n,L_m]_{(m-n)}= q^{-(n+m+5)}\Bigl(\,
q^{3(m-n)/4}G_{n/2}\{G_{n/2},G_{m/2}\}_{({m-n\over4})}G_{m/2} \nn\\
&&-q^{(m-n)/4}G_{n/2}G_{m/2}\{G_{n/2},G_{m/2}\}_{({m-n\over4})}
+\{G_{n/2},G_{m/2}\}_{({m-n\over4})}q^{(n-m)/4}G_{m/2}G_{n/2} \nn\\
&&-q^{3(n-m)/4}G_{m/2} \{G_{n/2},G_{m/2}\}_{({m-n\over4})} G_{n/2}
\,\Bigr) \nn \\
&&= q^{-(n+m+5)/2}(q^{(n-m)/4}+q^{(m-n)/4})\nn\\
&&\times\Bigl( 
q^{(m-n)/2}G_{n/2}[L_{n+m\over2},G_{m/2}]_{({m-n\over4})}
+q^{(n-m)/2}[L_{n+m\over2},G_{m/2}]_{({m-n\over4})}G_{n/2}\Bigr)\nn\\
&&= q^{-n-m-5/2}(q^{(n-m)/4}+q^{(m-n)/4})[{n-m\over4}]
\{G_{n/2},G_{m+n/2}\}_{({m-n\over2})}\ .
\eeqa
(ii) Assuming these constraints (in other words, no central extension 
for constraint algebras), we may find central extensions utilizing 
the Jacobi identities \eq{jacobGGG} and \eq{jacobFFF} 
\beq
\{G_r,G_s\}_{({s-r\over2})} = 
       q^{5/2+r+s} (q^{s-r\over2}+q^{r-s\over2})L_{r+s} 
+ f_1(r)\lambda\delta_{r+s,0} \ ,
\eeq
\beq
[F_n,F_m]_{(m-n)}= f_2(n)\lambda^2\delta_{n+m,0} \ ,
\eeq
where we leave the coefficients $f_i$ undetermined. These forms 
of central terms are consistent with Eqs.\eq{jacobLGG} and \eq{jacobLFF} 
as well. 

Secondly, let us decompose the above set of algebras into a $N=2$ 
algebra. The decomposition may be performed by 
\beq
G_r = G_r^- + G_r^+ \ ,
\eeq
with the two operators 
\beq
G_r^- = \lambda^{-1/2}x^{r+1/2}\delt, \quad 
G_r^+ = -\lambda^{-1/2}x^{r+1/2}\theta\delx \ . 
\eeq
The centerless $N=2$ CZ algebra is therefore 
\beq
\{G_r^+,G_s^- \} = q^{5/2+r+s} L_{r+s} + 
               q^{3/2+{r-s\over2}} [{r-s\over2}]F_{r-s}\ ,
\eeq
\beq
\{G_r^+,G_s^+\} = \{G_r^-,G_s^-\} = 0 \ ,
\eeq
\beq
[L_n, G_r^{\pm}]_{(r-{n\over2})} = 
              q^{-n}[{n\over2}-r] G_{n+r}^{\pm}\ ,
\eeq          
\beq
[F_n, G_r^+]_{(a,b)} = q^{n+2+a}\lambda G_{n+r}^+\ ,  \qquad
[F_n, G_r^-]_{(a,b)} = -q^{n+2r+1+b}\lambda G_{n+r}^-\ . 
\eeq
where $a$ and $b$ are ambiguous constants, and 
\beq
[A,B]_{(a,b)} = q^a AB- q^b BA\ .
\eeq 

To close this discussion, we finally put a brief comment on a 
coproduct structure. Originally, all these algebras are not 
defined in an abstract sense, but obtained as differential 
operator algebras acting on functions of quantum space variables. 
In the similar sense, we can derive some coproducts as 
Leipnitz rules for those differential operators, acting on 
a quantum superspace tensor bases $V^1_{mn}\otimes V^2_{mn}$ 
where $V^1_{mn}= x^{m}\theta^{n}$ etc.  For example, 
the operation rules (in other words, Leipnitz rules) 
\beqa
&& \delx x^n = q^{-2n} x^n \delx + q^{-n+1} [n] x^{n-1}\ ,\\
&& \lambda x^n = q^{-2n}x^n\lambda \ ,
\eeqa         
can be re-cast in the following forms of coproduct 
\beqa
\Delta(\delx)&=& \delx\otimes1  +  \lambda\otimes\delx  \ ,\\
\Delta(\lambda) &=& \lambda\otimes\lambda \ .
\eeqa
Similarly, we have 
\beqa
\Delta(L_n) &=&  L_n\otimes 1 + \lambda\otimes L_n \ , \label{copL}\\
\Delta(F_n) &=&  F_n\otimes 1 + \lambda\otimes F_n \ ,
\eeqa
and so forth. It is not so clear whether or not a fermionic 
operator (such as $\delt$ and $G_r$) can be organized into a 
unique coproduct expression. 
The relation \eq{newLn} would seem to suggest the structure 
\beq
\Delta(G_r)= G_r\otimes 1 + \lambda^{1/2}\otimes G_r \ .\label{copG}
\eeq
A check whether this works or not might be the following. 
Notice that Eqs.\eq{copL} and \eq{copG} for $n=0$, $r=\pm{1\over2}$ 
are the same coproduct structure as the one of the following 
Woronowitz type $q$-$osp(1,2)$ 
\beqa
&&{\tilde G}_{-1/2}{\tilde G}_{1/2}
     + q^{-1/2}{\tilde G}_{1/2}{\tilde G}_{-1/2}
     = \alpha(q){\tilde L}_0\ ,\nn \\
&&{\tilde L}_0{\tilde G}_{-1/2} 
     - q{\tilde G}_{-1/2}{\tilde L}_0 
     = q \gamma(q){\tilde G}_{-1/2}\ , \label{woro}\\
&&{\tilde L}_0{\tilde G}_{1/2} 
     - q^{-1}{\tilde G}_{1/2}{\tilde L}_0 
     = - \gamma(q){\tilde G}_{1/2}\ . \nn
\eeqa
The question then becomes whether or not a map between this 
algebra and ours keeps the coproduct structure unchanged. 
If one finds a mapping relation of our algebra into the 
standard $q$-$osp(1,2)$ algebra \cite{osp12}
\beq
[H,\; V_{\pm}] = \pm {1 \over 2} V_{\pm}, \qquad
\{V_+, V_-\} = - {1 \over 4} {p^{2H} - p^{-2H} \over p - p^{-1}}, 
\label{stand}
\eeq
our algebra will be transformed into the algebra \eq{woro} through 
the map 
\beqa
 & & {\tilde L}_0 = q \gamma {q^{2H}-1 \over q-1},\nn \\
 & & {\tilde G}_{-1/2} = 2\sqrt{q\alpha\gamma} V_+ q^{H/2}, \quad
     {\tilde G}_{1/2} = 2\sqrt{q\alpha\gamma} q^{H/2} V_-, \nn \\
 & & p = q^{1/2}\ .                                \nn
\eeqa 
It is known that the coproduct structure for \eq{stand} can be 
mapped into the one for \eq{woro}, and we expect that 
the problem would be solved in the similar way.

\vspace{5mm}
\noindent
{\bf Acknowledgement}
\vspace{5mm}
\noindent

The author would like to thank N. Aizawa and T. Kobayashi for discussions. 

%

\end{document}